\documentclass[9pt,twocolumn,twoside]{osajnl}

\journal{josab} 

\setboolean{shortarticle}{false}

\def\U#1{{%
\def\O{\mbox{O}}
\def\u{\mbox{u}}
\mathcode`\u=\mu
\mathcode`\O=\Omega
\mathrm{#1}}}
\def\Re{\mathop{\mathrm{Re}}}
\def\Im{\mathop{\mathrm{Im}}}
\def\ii{{\mathrm{i}}}
\def\dd{{\mathrm{d}}}
\def\sub#1{_{\scriptsize\mbox{#1}}}
\def\degree{\mbox{$^\circ$}}

\title{Tunable group delay in a doubly resonant metasurface composed of two dissimilar split-ring resonators}

\author[1,*]{Yasuhiro Tamayama}
\author[1]{Yutaro Kida}

\affil[1]{Department of Electrical, Electronics and
Information Engineering, Nagaoka University of
Technology, 1603-1 Kamitomioka, Nagaoka, Niigata 940-2188, Japan}

\affil[*]{Corresponding author: tamayama@vos.nagaokaut.ac.jp}

\doi{\url{https://doi.org/10.1364/JOSAB.36.002694}}

\begin{abstract}
We develop a method to control the group delay of electromagnetic waves continuously using a doubly resonant metasurface. The method is based on the dependences of i) the group velocity in a medium featuring two resonance lines on the resonance linewidths and ii) the resonance linewidth of a metasurface composed of split-ring resonators on an incidence angle of electromagnetic wave. To verify this method for group-delay control, we design a terahertz metasurface composed of two split-ring resonators with different resonance frequencies and numerically analyze the transmission characteristic of the metasurface. Double resonance lines are observed for oblique incidence and the resonance transmission dips become deeper and broader with increasing the incidence angle. The group delay at around the center frequency of the double resonance lines is found to vary in the range about from $0\,\U{s}$ to 20 times the period of the incident wave with the incidence angle. In contrast with a previously reported method for variable control of group delay using electromagnetically-induced-transparency-like metamaterials, a high transmittance is achieved for a small group delay condition.
\end{abstract}

\setboolean{displaycopyright}{true}

\begin{document}

\maketitle

\section{Introduction}

Metamaterials enable us to develop methods for controlling electromagnetic waves that cannot be achieved using only naturally occurring media. 
One of the most interesting topics in the field of metamaterials is the control of the group delay (group velocity) using metamaterials that mimic electromagnetically-induced-transparency (EIT).
Electromagnetically-induced-transparency  is a quantum interference phenomenon \cite{fleischhauer_05_rmp} and has been used for realizing very low group velocities \cite{hau_99_nature} and storage of optical pulses \cite{liu_01_nature,phillips_01_prl}.
As complicated systems are necessary to realize EIT, methods for mimicking EIT using metamaterials and other classical systems are under intensive study.
Such metamaterials are designed based on a mechanical model or an electrical circuit model of EIT \cite{alzar_02_ajp}.
A unit cell of EIT-like metamaterials is composed of two dissimilar resonators coupled with each other \cite{zhang_s_08_prl,tassin_09_prl,liu_n_09_nat_mater,lu_10_opex,tamayama_12_prb, rana_18_prappl,bagci_18_jap,xu_z_19_prb}. 
One resonator has a low $Q$ and couples to incident electromagnetic waves. The other has a high $Q$ and cannot be excited directly by the incident waves. 
Metamaterials composed of two radiative resonators coupled with each other have also been shown to exhibit EIT-like responses \cite{jin_x-r_12_jap,tamayama_14_prb,tamayama_15_prb, hokmabadi_15_sci_rep}.
In addition to these fundamental demonstrations of EIT-like metamaterials, active control of electromagnetic responses \cite{tamayama_10_prb,kurter_11_prl,gu_12_nat_comm, miyamaru_14_sci_rep,xu_q_16_ol,pitchappa_16_apl,yahiaoui_17_apl,yahiaoui_18_prb} and enhancement of nonlinear phenomena \cite{sun_y_13_apl,bai_z_15_sci_rep,tamayama_17_jap, tamayama_18_apl,chen_19_apa} in EIT-like metamaterials have also been investigated.
Furthermore, storage and retrieval of electromagnetic pulse waves have been demonstrated using EIT-like metamaterials \cite{nakanishi_13_prb,nakanishi_18_apl} as demonstrated using the original EIT.

The original EIT in quantum systems can be used for dynamic and continuous control of the group velocity as well as storage of an optical pulse.
Therefore, it is natural to consider that EIT-like metamaterials should also be used for continuous control of the group delay. 
In the original EIT, the group velocity for the probe light is controlled by varying the incident power of the pump light. 
This incident pump power corresponds to the coupling strength between the low-$Q$ and high-$Q$ resonators in EIT-like metamaterials. 
The coupling strength depends on the physical distance between the low-$Q$ and high-$Q$ resonators, and therefore, the distance between the two resonators needs to be changed to control the group delay in EIT-like metamaterials \cite{zhang_s_08_prl,liu_n_09_nat_mater, rana_18_prappl,bagci_18_jap}.
Although this method for controlling group delay is quite analogous to that in the original EIT, microelectromechanical systems (MEMS) or other reconfigurable systems are necessary for variable control of the group delay. 
Alternatively, the group delay can be controlled by varying the loss in the high-$Q$ resonator. 
This idea for controlling group delay is different from that in the original EIT. 
Although varying the loss in the high-$Q$ resonator is relatively easy \cite{kurter_11_prl,gu_12_nat_comm,yahiaoui_17_apl}, the attenuation of electromagnetic waves increases with losses in the high-$Q$ resonator. 
While EIT-like metamaterials are very useful in storage and retrieval of electromagnetic waves, the above aspects imply that they may not be the best media for continuous control of the group delay. 
Specifically, where only a continuous control of the group delay is the focus, metamaterials have only to have similar transmission spectra to EIT and do not have to exhibit EIT-like microscopic responses. 

In this study, we develop a method for continuous control of the group delay using a metasurface with an EIT-like transmission spectrum that does not exhibit an EIT-like microscopic response.
We focus on a medium with double resonance lines as a medium that has a similar transmission spectrum to EIT.
At the center frequency of the double resonance lines, a high transmittance and a low group velocity are achieved.
To control the group velocity at the center frequency of the double resonance lines, the frequency difference between the two resonance lines or the resonance linewidths need to be varied.
Because the resonance frequency of a metamaterial depends on the structural parameters and the medium parameters such as permittivity, MEMS or active elements must be introduced to vary the resonance frequency. In contrast, 
as the resonance linewidth of a metamaterial depends on the relationship between the configuration of the constituent meta-atoms and the propagation direction of the incident electromagnetic wave \cite{tao_09_prl}, the resonance linewidth can be varied by simply varying the incidence angle on the metamaterial. 
We describe the theory governing the control of the group delay based on this idea and present a numerical verification of the theory in the following.
Also, we compare characteristics of group delay control using EIT-like metamaterials and the present metasurface.

\section{Theory}

First, we describe a theory for continuous control of the group delay using a medium with double resonance lines.
We assume that the response of the medium is modeled as doubly resonant Lorentz oscillators and that the refractive index of the medium is given by
\begin{equation}
n(\omega) = 1 
- \gamma_1 \frac{\alpha}{\omega^2 + \ii \gamma_1 \omega - \omega_1^2} 
- \gamma_2 \frac{\alpha}{\omega^2 + \ii \gamma_2 \omega - \omega_2^2} , \label{eq:2-10}
\end{equation}
where $\omega$ is the angular frequency of the incident electromagnetic wave, $\omega_{1,2}$ are the resonance angular frequencies, $\gamma_{1,2}$ the resonance linewidths, and $\alpha$ is a constant. 
While the second/third term gives the electric susceptibility in the Lorentz oscillator model, we assume the refractive index is written as in \eqref{eq:2-10} for simplicity. 
The extra $\gamma_{1,2}$ in the second/third term reflects the property that the susceptibility increases with the resonance linewidth, the reason for which is described in the next paragraph. 
We now calculate the group index at $\omega = \omega_0 = (\omega_1 + \omega_2)/2$ based on \eqref{eq:2-10}. 
The group index is given by $n\sub{g} = n + \omega \Re{(\dd n / \dd \omega)}$ and may be approximated by $n\sub{g} \approx \omega \Re{(\dd n / \dd \omega)}$ at around $\omega = \omega_0$ because of the steep dispersion near the center frequency of the double resonance lines. Assuming that the difference between $\omega_1$ and $\omega_2$ is small and that $\gamma_1 = \gamma_2 = \gamma_0$, the group index at $\omega = \omega_0$ is further approximated yielding
\begin{equation}
n\sub{g} \approx \omega_0
\Re{\left( \left. \frac{\dd n}{\dd \omega} \right|_{\omega=\omega_0} \right)}
\approx 4\alpha\gamma_0 \frac{\delta^2 - \gamma_0^2}{(\delta^2 + \gamma_0^2)^2} ,
\label{eq:2-20}
\end{equation}
where $\delta = \omega_2 - \omega_1$. 
As $\gamma_0$ increases from 0, $n\sub{g}$ first increases, reaches a maximum value $(n\sub{g})\sub{max}$ at $\gamma_0 = \sqrt{3-2\sqrt{2}}\delta$ ($\approx 0.41\delta$), and then decreases. 
If $\gamma_0$ further increases and exceeds $\delta$, $n\sub{g}$ becomes negative. 
This implies that double resonance lines are observed for $\gamma_0 < \delta$ whereas only a single resonance line is observed for $\gamma_0 > \delta$ because of a large overlap of the resonance lines. 
Assuming that the refractive index at $\omega = \omega_0$ is $n_0$ independent of $\gamma_0$, the group index at $\omega = \omega_0$ can be controlled in the range from $n_0$ to $(n\sub{g})\sub{max}$ by varying $\gamma_0$ in the range from $0$ to $\sqrt{3-2\sqrt{2}}\delta$ or from $\sqrt{3-2\sqrt{2}}\delta$ to about $\delta$.
Because the value of $\Im{[n(\omega_0)]}$, which corresponds to the transmission loss, increases with $\gamma_0$, $\gamma_0$ must vary in the range from 0 to $\sqrt{3-2\sqrt{2}}\delta$ for group delay control with low transmission loss.

\begin{figure}[tb]
\begin{center}
\includegraphics[scale=0.5]{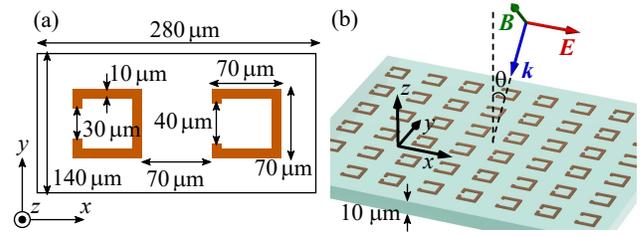}
\caption{(a) Geometry of the unit structure of the metasurface and (b) relationship between the incident electromagnetic fields and SRRs. }
\label{fig:geometry}
\end{center}
\end{figure}

The above method for controlling the group delay is realized using a metasurface with a unit structure shown in Fig.~\ref{fig:geometry}(a).
This unit structure consists of two split-ring resonators (SRRs) with different resonance frequencies. 
Here, an electromagnetic wave is assumed to be incident on the metasurface as shown in Fig.~\ref{fig:geometry}(b). 
In this configuration, the incident electric field, which is along the $x$ direction for all angles of incidence, cannot excite the inductor--capacitor resonance of the SRRs; only the incident magnetic field can excite this resonance. 
For oblique incidence, the SRRs are excited and two resonance dips appear in the transmission spectrum at the resonance frequencies of the SRRs. 
As the incidence angle $\theta$ increases, the incident magnetic flux that goes through the SRRs increases and the electromotive force induced in the SRRs increases. 
Furthermore, the values of the radiation pattern of the induced magnetic dipole in the SRRs for the propagation directions of the reflection and transmission waves increase with $\theta$, which implies that the radiation loss, i.e., the resonance linewidth, of the metasurface increases with $\theta$.
Therefore, the group delay at around the center frequency of the double resonance lines can be controlled by the incidence angle on the metasurface. 
To take into account the dependences of the electromotive force and of the resonance linewidth on the incidence angle, $\gamma_{1,2}$ is factored with the susceptibility derived from the Lorentz oscillator model in \eqref{eq:2-10}.

\section{Simulation and discussion}

We numerically analyzed the transmission characteristic of the metasurface shown in Fig.~\ref{fig:geometry} using a commercial finite-element software COMSOL Multiphysics to verify that the group delay in the metasurface can be controlled by varying the incidence angle of the electromagnetic wave.
The SRRs were assumed to be made of perfect electric conductor with vanishing thickness to reduce memory requirements in the simulation. 
This assumption creates no significant influence on the results in the low-frequency regions such as the microwave and terahertz frequency regions \cite{sasaki_18_apa}. 
The substrate was assumed to be made of Benzocyclobutene (BCB). 
The thickness and permittivity of the substrate were set to be $10\,\mu\U{m}$ and $2.56+\ii 0.01792$, respectively.
The reason why BCB was adopted is that, with high refractive index, the magnetic response of the SRRs remains weak even for grazing incidence because of refraction. 
In addition, preliminary simulations showed that the magnetic response of SRRs becomes stronger for thinner substrates. 
Therefore, the substrate should be thin and made of a material with a low refractive index. The refractive index of BCB is 1.6 and
metamaterial structures can be fabricated on BCB films with thickness of about $10\,\mu\U{m}$ \cite{paul_08_opex}.
We thus decided that BCB is a suitable material to use as a substrate for the present metasurface. 
We add that polyimide, which has a slightly higher refractive index and loss than BCB, may also be used as the substrate \cite{tao_08_jphysd}.
In this simulation, periodic boundary conditions were applied to the $x$ and $y$ directions and perfectly matched layer boundary condition was applied to the $z$ direction.

\begin{figure}[tb]
\begin{center}
\includegraphics[scale=0.68]{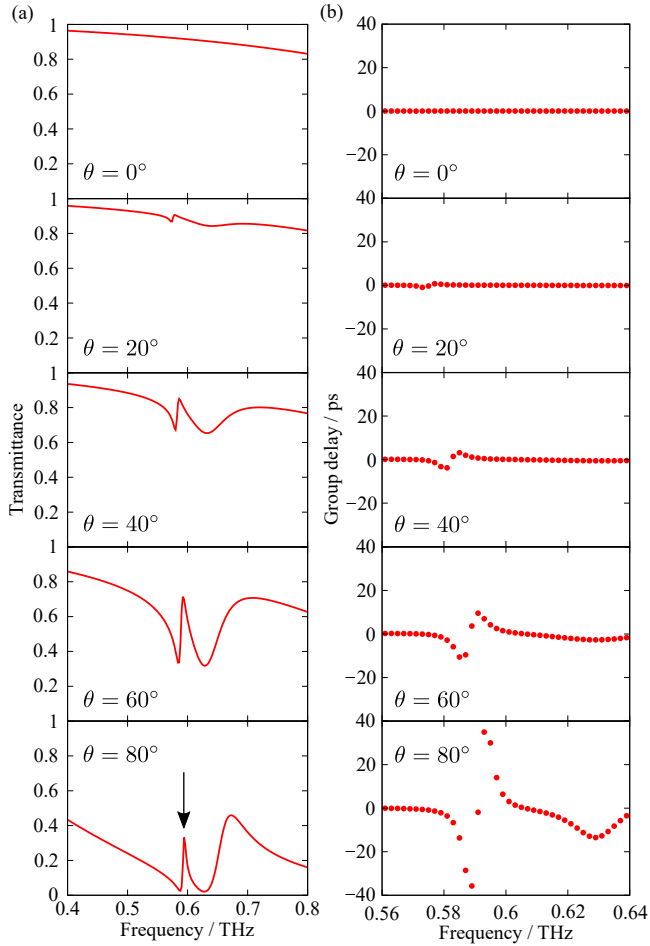}
\caption{Frequency dependences of (a) transmittance and (b) group delay of the metasurface shown in Fig.~\ref{fig:geometry} for $\theta=0\degree$, $20\degree$, $40\degree$, $60\degree$, and $80\degree$. Note that the horizontal scale in (a) and (b) differ.} 
\label{fig:trans_uni}
\end{center}
\end{figure}

The numerically calculated transmission spectra for various angles of incidence are shown in Fig.~\ref{fig:trans_uni}(a). An almost flat spectrum is observed for normal incidence and two resonance dips are observed at around $0.58\,\U{THz}$ and $0.63\,\U{THz}$ for oblique incidence.
As the incidence angle increases, the resonance dips deepen and broaden.
This implies that a medium with refractive index given by \eqref{eq:2-10} is qualitatively realized and that $\gamma_0$ varies significantly by varying the incidence angle.
We add that the transmittance decreases toward the higher frequency region even for normal incidence because of the half-wavelength resonance of the SRRs induced by the incident electric field. 
Figure \ref{fig:trans_uni}(b) shows 
the frequency dependence of the group delay for various angles of incidence. The maximum value increases with the incidence angle. 
The qualitative reason for this monotonic increase is that the difference in the resonance frequencies of the two SRRs is relatively large and $\gamma_0$ does not exceed $\sqrt{3-2\sqrt{2}}\delta$ for $\theta \leq 80\degree$.
From these results, the group delay at around $\omega = \omega_0$ is found to be controllable in the range about from $0\,\U{s}$ to 20 times the period of the incident electromagnetic wave by varying the incidence angle in the metasurface designed based on \eqref{eq:2-10} with realistic parameters.

\begin{figure}[tb]
\begin{center}
\includegraphics[scale=0.4]{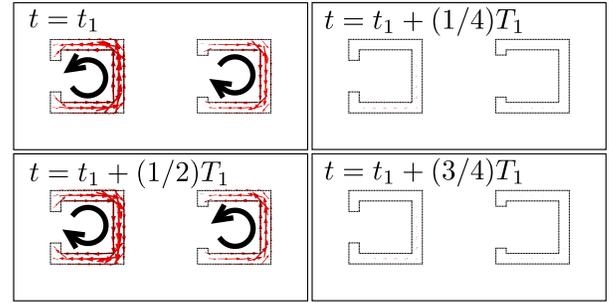}
\caption{Snapshots of the current distribution in the metasurface shown in Fig.~\ref{fig:geometry}
at the incidence frequency of $0.594\,\U{THz}$ for $\theta = 80\degree$, which is indicated by the arrow in Fig.~\ref{fig:trans_uni},
where $t_1$ is a certain time and $T_1$ is the oscillation period.
The black arrows are solely a visual guide and indicate the direction of current flow in the SRRs.}
\label{fig:current_uni}
\end{center}
\end{figure}

We discuss next the influence of bianisotropy of the SRRs \cite{marques_02_prb,gay-balmaz_02_jap, katsarakis_04_apl,garcia_05_jap,zhou_05_prl,koschny_05_prb_2} on the transmission characteristic of the metasurface. 
In this study, single-gap SRRs, which exhibit bianisotropy, were used as meta-atoms. 
To realize as closely as possible metasurfaces with characteristics given by \eqref{eq:2-10}, double-gap SRRs may seem to be more suitable for meta-atoms than single-gap SRRs because double-gap SRRs do not exhibit bianisotropy. 
However, single-gap SRRs were adopted in this study.
This is because through preliminary numerical analyses the magnetic response of a metasurface composed of single-gap SRRs was found to be stronger than that of double-gap SRRs. 
That is, the resonance transmission dip for a metasurface composed of single-gap SRRs was deeper than that of double-gap SRRs when the incidence angles for the two configurations were the same.
One may consider that an unnecessary polarization conversion may be induced in a metasurface composed of single-gap SRRs because of the bianisotropy. 
Actually, the influence of bianisotropy vanishes at around $\omega = \omega_0$ for the configuration shown in Fig.~\ref{fig:geometry}.
This can be understood from the current distribution in the SRRs at $\omega = \omega_0$.
Figure \ref{fig:current_uni} shows the current distribution at the incident frequency of $0.594\,\U{THz}$ for $\theta=80\degree$. 
An electric dipole as well as a magnetic dipole are induced by the incident magnetic field in each SRR because of the bianisotropy. 
The oscillations in the current flows in the two SRRs are in anti-phase at this frequency. 
Roughly speaking, this is because the resonance frequency of one SRR is lower than $\omega_0$ and that of the other SRR is higher than $\omega_0$. (The oscillation phase difference in the two SRRs is discussed in more detail in the next paragraph.)
Thus, the oscillations of the electric dipoles induced in the two SRRs are in anti-phase, which results in a cancellation of the total electric dipole.
Therefore, at around $\omega = \omega_0$, the influence of the bianisotropy vanishes and the group delay control without an unnecessary polarization conversion can be realized.

\begin{figure}[tb]
\begin{center}
\includegraphics[scale=0.68]{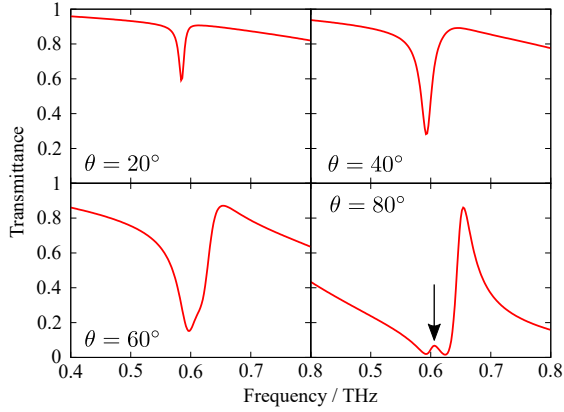}
\caption{Transmission spectra of the metasurface for $\theta = 20\degree$, $40\degree$, $60\degree$, and $80\degree$ when one of the SRRs shown in Fig.~\ref{fig:geometry}(a) is rotated $180\degree$ about the $z$-axis. }
\label{fig:trans_sym}
\end{center}
\end{figure}

We also discuss the influence of coupling between these two SRRs.
The transmission spectra shown in Fig.~\ref{fig:trans_uni}(a) are slightly asymmetric about the center frequency of the double resonance lines; that is, the spectra are a little bit different from spectra with double Lorentzian resonance lines given by \eqref{eq:2-10}.
This implies that a coupling exists between the two SRRs. 
This coupling is clearly confirmed when either of the SRRs in Fig.~\ref{fig:geometry}(a) is rotated $180\degree$ about the $z$-axis.
The transmission spectrum of such a metasurface is shown in Fig.~\ref{fig:trans_sym}. 
The resonance dip deepen and broaden with the increase in the incidence angle but only a single resonance line is observed. 

\begin{figure}[tb]
\begin{center}
\includegraphics[scale=0.4]{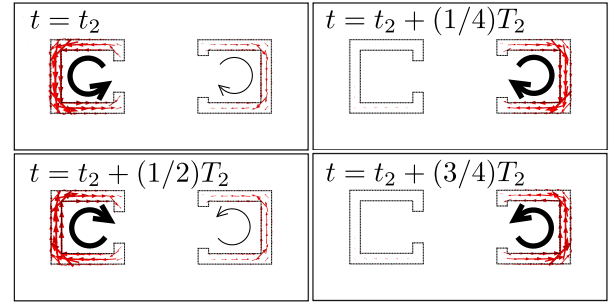}
\caption{Snapshots of the current distribution in the metasurface that corresponds to Fig.~\ref{fig:trans_sym} at $0.606\,\U{THz}$ for $\theta = 80\degree$, which is indicated by the arrow in Fig.~\ref{fig:trans_sym}, where $t_2$ is a certain time and $T_2$ is the oscillation period.  The black arrows are a visual guide indicating the direction and intensity of current flow in the SRRs.}
\label{fig:current_sym}
\end{center}
\end{figure}

The difference between the spectra shown in Figs.~\ref{fig:trans_uni}(a) and \ref{fig:trans_sym} never occurs without the coupling between the two SRRs. 
The coupling is unnecessary in the theory based on \eqref{eq:2-10} but the coupling is unavoidable because the polarizations of the radiation from the two SRRs are the same. 
The influence of the coupling can be understood based on a coupled resonator model \cite{tamayama_14_prb}.
Assuming that the two SRRs are modeled as inductor--capacitor resonant circuits coupled via a mutual impedance $Z\sub{M}$, the ratio of the stored charges $q_1$ and $q_2$ in the two capacitors at $\omega = \omega_0$ is found to be
\begin{equation}
\frac{q_1}{q_2} \approx \frac{-2 ( \omega_0 - \omega_1) 
+ \ii [ \gamma_0 - (Z\sub{M} / L)] }{2 ( \omega_0 - \omega_1)
+ \ii [ \gamma_0 - (Z\sub{M} / L)] }, 
\label{eq:3-10}
\end{equation}
assuming the inductances of the two SRRs are the same. 
When the coupling $Z\sub{M}$ and the loss $\gamma_0$ vanish, the value of \eqref{eq:3-10} equals $-1$; that is, the current oscillations in the two SRRs are in anti-phase. 
This implies that the radiations from the two SRRs cancel each other out and that perfect transmission occurs at $\omega = \omega_0$. 
In this instance, double resonance lines are clearly observed. 
If the coupling and/or the loss is present, the oscillation phase difference can become different from $180\degree$. 
When only the loss is present, the oscillation phase difference becomes smaller than $180\degree$ and the cancellation of radiation is incomplete. 
A decrease in the transmittance results at $\omega = \omega_0$ and eventually the double resonance lines changes into a single resonance line. 
If the coupling is also present, the oscillation phase difference becomes close to (far from) $180\degree$ depending on the argument of $Z\sub{M}$.
In the present situation, the center frequency of the transmission peak in Fig.~\ref{fig:trans_uni}(a) is not so far from the average of the resonance frequencies of the two SRRs, which implies that $\Im{(Z\sub{M})}$ is small \cite{tamayama_14_prb}.
Therefore, we here on assume for simplicity that $\Im{(Z\sub{M})} = 0$. 
When $\Re{(Z\sub{M})} > 0$, the effect of the radiation loss $\gamma_0$ is suppressed by the coupling $Z\sub{M}$ and the oscillation phase difference approaches $180\degree$ compared with that for $Z\sub{M}=0$. 
That is, the double resonance lines are observed more clearly because of the coupling. 
When $\Re{(Z\sub{M})} < 0$, the imaginary parts of the numerator and denominator in \eqref{eq:3-10} increase and the oscillation phase difference becomes far from $180\degree$ compared with zero-coupling instances.
This implies that the radiations from the SRRs are enhanced because of the coupling and that the transmittance at $\omega = \omega_0$ decreases. 
In this instance, the two resonance lines overlap more obscuring the two distinct resonance transmission dips. 
To confirm the influence of the coupling, we compare the numerically calculated current distributions in the two different metasurfaces. 
As described in the previous paragraph, the oscillations of the current flows in the SRRs shown in Fig.~\ref{fig:current_uni} are in anti-phase, implying that $\Re{(Z\sub{M})} > 0$. 
This is the reason why the double resonance lines are clearly observed in the metasurface shown in Fig.~\ref{fig:geometry}. 
Figure \ref{fig:current_sym} shows the current distribution in the metasurface that corresponds to Fig.~\ref{fig:trans_sym}. 
The oscillation phase difference of the induced currents in the SRRs is found to be quite different from $180\degree$ and almost equal to $90\degree$. 
This implies that $\Re{(Z\sub{M})} < 0$, and therefore, only a single resonance line is observed in Fig.~\ref{fig:trans_sym}.
We add that the difference in the coupling phase between the two instances is caused by the bianisotropy of the SRRs.
The above discussion shows that meta-atoms need to be arranged so that $\Re{(Z\sub{M})}$ becomes positive to realize the group-delay control described in this study.

Note that the variable control of the group delay by the incidence angle of electromagnetic wave has already been demonstrated in EIT-like metasurfaces in several studies \cite{lu_10_opex,tamayama_12_prb,jin_x-r_12_jap}.
In these EIT-like metasurfaces, the coupling strength between the low-$Q$ and high-$Q$ resonance modes depends on the gradient of the electromagnetic fields in the direction parallel to the plane of metasurface.
As the incidence angle increases, the coupling strength increases and the transmission spectrum gradually changes from a Lorentz-type spectrum to an EIT-like spectrum. 
The group delay at the transmission peak frequency first increases, reaches a maximum at a certain angle of incidence, and then decreases with increasing the incidence angle, which is similar to the characteristic of the metasurface in this study. 
The transmittance at the transmission peak frequency increases with the incidence angle, and hence, there exist small group delay conditions with high and low transmittances. 
Obviously, a small group delay with a high transmittance is preferable to that with a low transmittance.  The former can be achieved in principle if the coupling becomes very strong, 
however, realizing such a strong coupling is difficult even for a large angle of incidence. 
Therefore, the transmittance for a small group delay condition is very low in these EIT-like metasurfaces. 
This property is not suitable in variable control of the group delay with low loss. 
In contrast, with the present metasurface, the transmission spectrum gradually changes from a flat spectrum to an EIT-like spectrum as the incidence angle increases. 
The characteristic that the group delay becomes maximum at a certain angle of incidence is the same as that in the above EIT-like metasurfaces. 
The difference is that a small group delay with a high transmittance can be achieved at around normal incidence in the present metasurface.
This property is preferable for low-loss tuning of the group delay. 

\section{Conclusion}

We investigated variable control of the group delay using a metasurface featuring double resonance lines.
The method of control is based on a characteristic that the resonance linewidth of a SRR metasurface depends on the relationship between the configuration of SRRs and the propagation direction of the incident electromagnetic wave.
A metasurface composed of two SRRs with different resonance frequencies was designed to realize incident-angle-dependent double resonance lines. 
Through numerical simulation, double resonance lines were observed for oblique incidence; resonance strengths and resonance linewidths of the SRRs were seen to increase with the incidence angle. 
The group delay near the center frequency of the double resonance lines was found to increase with the incidence angle. 
The range in variation of the group delay was about from $0\,\U{s}$ to 20 times the period of the incident electromagnetic wave. 
In addition, a high transmittance was achieved for a small group delay condition in the present metasurface in contrast with previously developed EIT-like metamaterials. 
Note that there still remains an issue that the transmittance decreases with increasing the group delay and becomes a low value for the maximum group delay condition. If this issue is solved in future studies, transmissive variable pulse delay lines with compact dimensions would be realized.

The microscopic response of the metasurface is quite different from EIT-like metamaterials, whereas the macroscopic transmission characteristic is similar to EIT. 
Using both of metamaterials with double resonance lines and EIT-like metamaterials as the situation demands would be important in the further development of methods for controlling electromagnetic pulse waves. 

\section*{Funding}

\mbox{JSPS} \mbox{KAKENHI} (\mbox{JP16H06086}).



\end{document}